\documentclass[
 reprint,
 amsmath,
 amssymb,
 aps,
]{revtex4-2}
\usepackage[utf8]{inputenc}
\usepackage{graphicx}
\usepackage{hyperref}
\usepackage{gensymb}
\usepackage{siunitx}
\newcommand{\figsize}{1.}

\begin{document}

\title{Measurement of the Ra$^+$ $7p$ $^{2}P_{3/2}$ state lifetime}
\author{M. Fan}
\affiliation{Department of Physics, University of California, Santa Barbara, California 93106, USA}
\author{C. A. Holliman}
\affiliation{Department of Physics, University of California, Santa Barbara, California 93106, USA}
\author{A. Contractor}
\affiliation{Department of Physics, University of California, Santa Barbara, California 93106, USA}
\author{C. Zhang}
\affiliation{Department of Physics, University of California, Santa Barbara, California 93106, USA}
\author{S. F. Gebretsadkan}
\affiliation{Department of Physics, University of California, Santa Barbara, California 93106, USA}
\author{A. M. Jayich}
\email{jayich@gmail.com}
\affiliation{Department of Physics, University of California, Santa Barbara, California 93106, USA}

\date{\today}

\begin{abstract}
    We report a measurement of the radium ion's $7p$ ${}^{2}P_{3/2}$ state lifetime, $\tau=4.78(3)$ ns. The measured lifetime is in good agreement with theoretical calculations, and will enable a determination of the differential scalar polarizability of the narrow linewidth $7s$ ${}^{2}S_{1/2}\rightarrow$ $6d$ ${}^{2}D_{5/2}$ optical clock transition. 
\end{abstract}
\maketitle

\section{Introduction}

The radium ion is a promising candidate for an optical clock \cite{Sahoo2007, Holliman2022}, which requires precise knowledge of the atom's properties. Measurements of excited state lifetimes and branching fractions can be used to calculate the dipole matrix elements, which are used to determine leading systematic effects of the optical clock such as the blackbody radiation shift and laser-induced ac Stark shifts \cite{Dube2013}. As the lifetimes of the $\mathrm{Ra}^+$ $7p$ states have not been measured, the current assessment of the $\mathrm{Ra}^+$ ion clock systematics relies on the theoretically calculated dipole transition matrix elements \cite{Sahoo2009}. This results in a fractional frequency uncertainty due to the blackbody radiation that is greater than $10^{-17}$ \cite{Holliman2022}. Measurements of the $7p$ state lifetimes and of a $7s\ {}^2S_{1/2}\rightarrow 6d\ {}^2D_{5/2}$ magic wavelength can reduce this uncertainty \cite{Barrett2019}.

Precision measurements of the $\mathrm{Ra}^+$ excited-state lifetimes can also be used to benchmark theoretical calculations of heavy atoms.  The only electronic state lifetime measurements of elements in the last row of the periodic table with sufficient precision to test theory (less than $1\%$ fractional uncertainty) have been done with Fr \cite{Zhao1997, Simsarian1998, Grossman2000, Aubin2003, Gomez2005}.  This lifetime measurement adds another benchmark for theoretical calculations of heavy elements.  Accurate theory in such systems is needed for a ${}^{229}\mathrm{Th}^{3+}$ nuclear clock \cite{Campbell2012} and for measuring parity non-conservation with Fr \cite{Gomez2006}.

In this work we measure the lifetime of the $7p$ ${}^{2}P_{3/2}$ state using a single laser-cooled ${}^{226}\mathrm{Ra}^{+}$ ion (nuclear spin $I=0$) by comparing absorptive and dispersive interactions due to off-resonant light. This technique was proposed in \cite{Gerritsma2008} and was used to measure the lifetimes of the Ca$^+$ $4p$ $^{2}P_{1/2}$ \cite{Hettrich2015} and $^{2}P_{3/2}$ states \cite{Meir2020}, and the Lu$^+$ $6s6p$ $^{3}P_{0}$ and $^{3}P_{1}$ states \cite{Arnold2019a}.

\section{Theory}

\begin{figure}
    \centering
    \includegraphics[width=\figsize\linewidth]{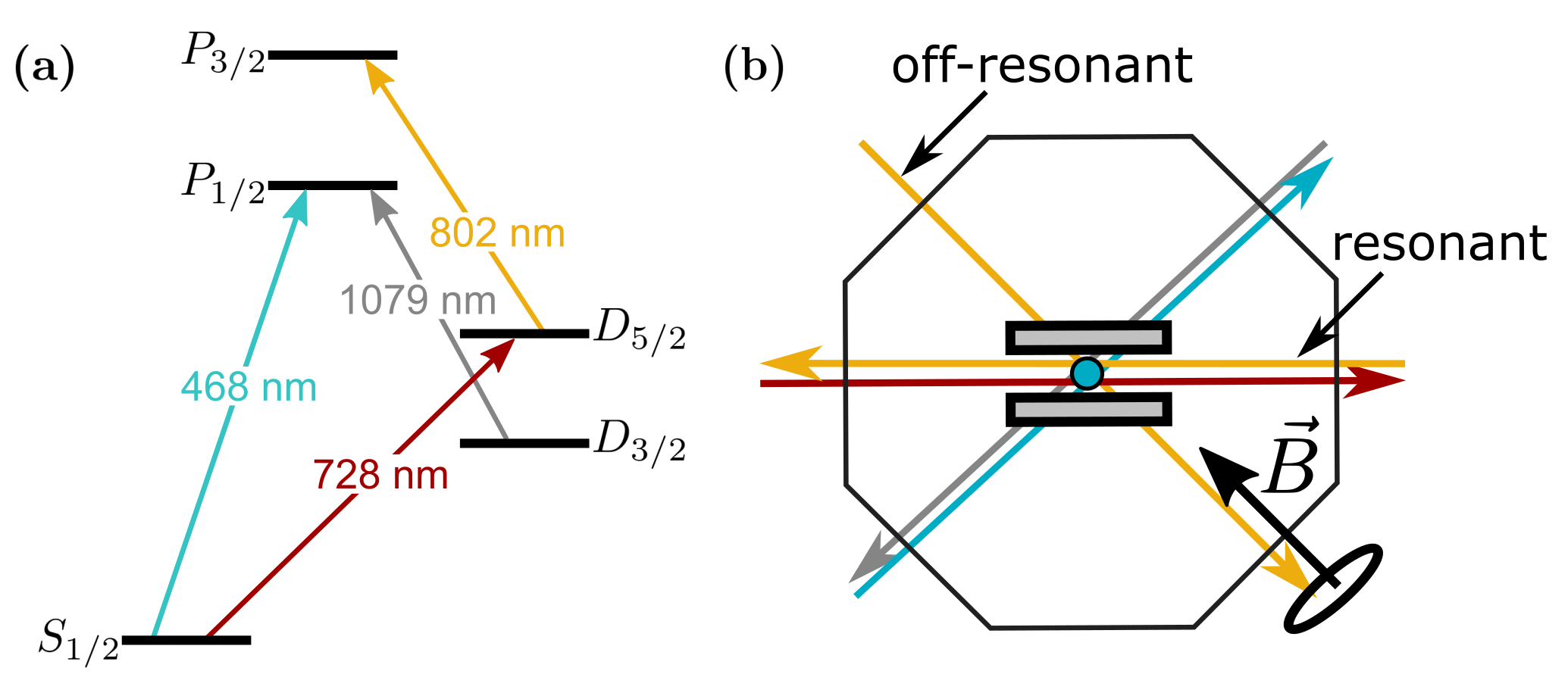}
    \caption{(a) $\mathrm{Ra}^+$ energy levels and transitions used in this work. (b) Geometry of the experimental setup highlighting the resonant and off-resonant 802-nm light. The $\mathrm{Ra}^+$ ion (cyan) is trapped in a linear Paul trap with the radial electrodes shown. The wavelength of lasers corresponds to colors of different transitions in (a).}
    \label{fig:trap-levels}
\end{figure}

The low-lying states of $\mathrm{Ra}^+$ are shown in Fig. \ref{fig:trap-levels}(a). We drive the $D_{5/2}\rightarrow P_{3/2}$ transition (802 nm) off-resonantly, which ac Stark shifts the $D_{5/2}$ state (a dispersive interaction) and drives population from the $D_{5/2}$ state to the $S_{1/2}$ and $D_{3/2}$ states (an absorptive interaction). The $P_{3/2}$ state lifetime is determined from the ratio between the dispersive and absorptive interactions.

In this work, the off-resonant light detuning is much smaller than the 802-nm transition frequency, so we can apply the rotating-wave approximation and consider only the corotating term \cite{Meir2020}. Moreover, the detuning is much larger than the natural linewidth of the 802 nm transition and the Rabi frequency of the off-resonant light, so contributions from the transition line shape are neglected (see Appendix \ref{subsec:line_shape}). Under these conditions, the ac Stark shift of a $D_{5/2}$ Zeeman sublevel due to the off-resonant 802-nm light is given by

\begin{equation}
    \delta_{\mathrm{S}}(m_D) = \sum_{m_P}\frac{\Omega(m_D, m_P)^2}{4\Delta(m_D, m_P)},
\label{eq:ac_stark_shift}
\end{equation}
where $m_D$ and $m_P$ are the magnetic quantum numbers of Zeeman sublevels of the $D_{5/2}$ and $P_{3/2}$ states, $\Omega(m_D, m_P)$ is the Rabi frequency for a transition from the $m_D$ sublevel to the $m_P$ sublevel, and $\Delta(m_D, m_P)$ is the detuning of the off-resonant light from the transition.

The pumping rate from the $D_{5/2}$ state to the $S_{1/2}$ and $D_{3/2}$ states due to the off-resonant 802-nm light is

\begin{equation}
    \mathcal{R}(m_D) = (A_{P_{3/2}\rightarrow S_{1/2}}+A_{P_{3/2}\rightarrow D_{3/2}})\sum_{m_P}\frac{\Omega(m_D, m_P)^2}{4\Delta(m_D, m_P)^2},
\label{eq:pumping}
\end{equation}
where $A_{P_{3/2}\rightarrow S_{1/2}}$ and $A_{P_{3/2}\rightarrow D_{3/2}}$ are the spontaneous decay rates of the $P_{3/2}$ state to the $S_{1/2}$ and $D_{3/2}$ states, respectively.

Given that the Zeeman splittings between transitions in the experiment (approximately $10$ MHz) are much smaller than the 802 nm light detuning (greater than 1.5 GHz), we can approximate $\Delta(m_D, m_P)\approx\Delta(m_D)$, where $\Delta(m_D)$ is the average detuning of all electric dipole allowed Zeeman transitions from the $m_D$ sublevel (see Appendix \ref{subsec:zeeman_splittings}). With this approximation, the detuning terms in Eqs. (\ref{eq:ac_stark_shift}) and (\ref{eq:pumping}) are independent of the summation over $m_P$, and the decay rates can be calculated from a ratio of the ac Stark shift and the pumping rate

\begin{equation}
    A_{P_{3/2}\rightarrow S_{1/2}}+A_{P_{3/2}\rightarrow D_{3/2}} = \Delta(m_D)\frac{\mathcal{R}(m_D)}{\delta_{\mathrm{S}}(m_D)}.
\label{eq:einstein_a}
\end{equation}

With Eq. (\ref{eq:einstein_a}) and the measured branching fraction of the $P_{3/2}$ state to the $D_{5/2}$ state $p=\SI{0.10759\pm0.00010}{}$ \cite{Fan2019a}, the $P_{3/2}$ state lifetime is

\begin{equation}
    \tau = \frac{1-p}{\Delta(m_D)}\frac{\delta_{\mathrm{S}}(m_D)}{\mathcal{R}(m_D)}.
\label{eq:lifetime}
\end{equation}

\section{Experiment}

We measure the ac Stark shift and the pumping rate using a single $^{226}$Ra$^+$ ion in a linear Paul trap with diagonal radial electrode separation $2r_0=6$ mm, axial electrode separation $2z_0=15$ mm, and radial rf frequency $\omega_{\mathrm{rf}}=2\pi\times993$ kHz, as described in \cite{Holliman2022}. The radial secular frequencies are $\omega_\mathrm{r}\sim2\pi\times110$ kHz and the axial secular frequency is $\omega_\mathrm{a}=2\pi\times37$ kHz.

\begin{figure*}
    \centering
    \includegraphics[width=\linewidth]{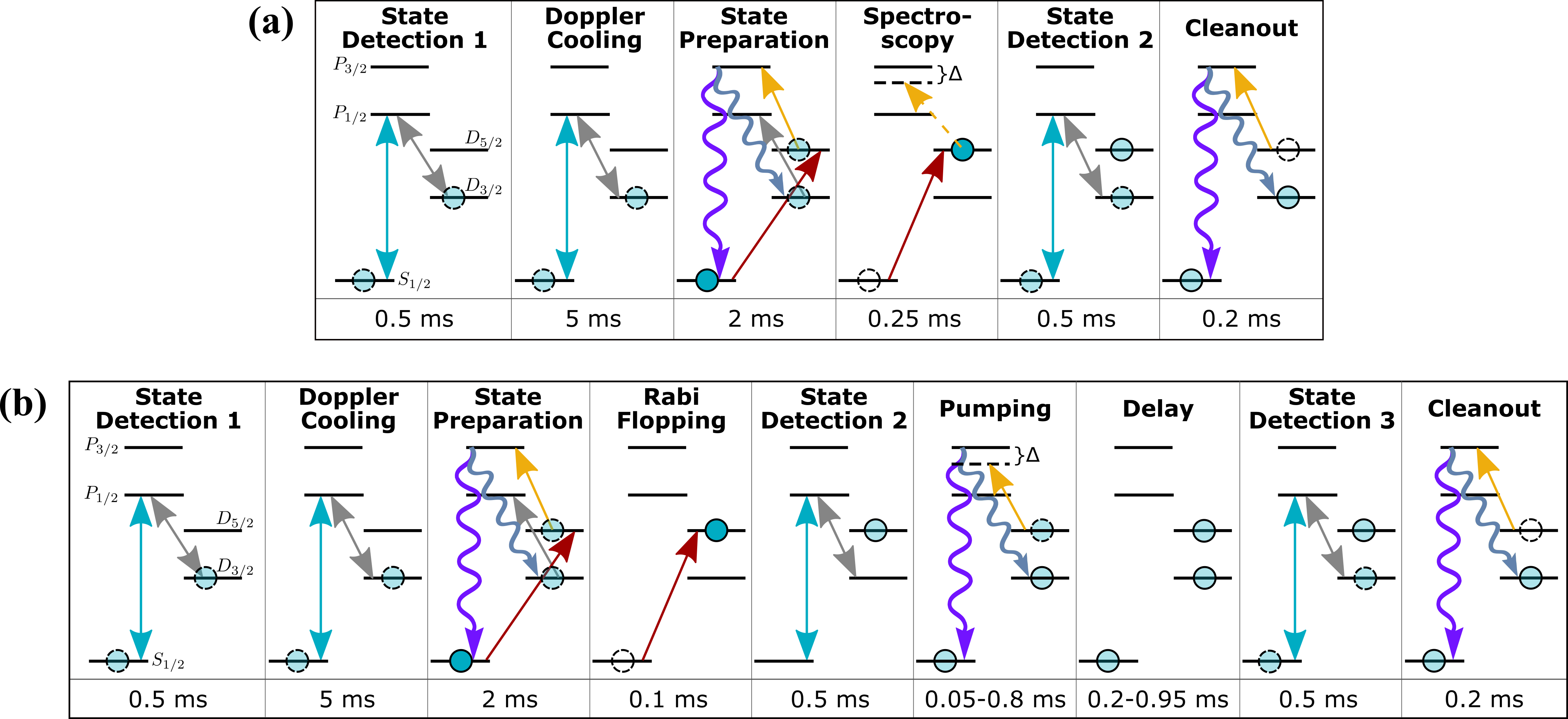}
    \caption{Pulse sequences for the (a) ac stark shift measurement and (b) pumping rate measurement. Squiggly lines depict $E1$ allowed decays, straight lines show optical pumping transitions, double-headed arrows indicate optical cycling transitions, and dashed arrows represent pulses that are not used for every repetition of the sequence. During the delay step in (b), no light is applied to the ion. The delay step maintains the same 1-ms interval between state detections 2 and 3 for different off-resonant light pulse durations in order to keep the effect of $D_{5/2}$ spontaneous decays constant.}
    \label{fig:sequence}
\end{figure*}

The experiment setup is shown in Fig. \ref{fig:trap-levels}(b). A static magnetic field of $\SI{5.737\pm0.002e-4}{\tesla}$ is applied to set the quantization axis. Five lasers are used to control the ion. The ion is Doppler cooled using 468- and 1079-nm light. To populate the $D_{5/2}$ state, a 728-nm laser drives the $S_{1/2}\rightarrow D_{5/2}$ electric quadrupole transition. The $D_{5/2}$ state is pumped back into the Doppler cooling cycle by a 802-nm laser that is resonant with the $D_{5/2}\rightarrow P_{3/2}$ transition. A second 802-nm laser provides the off-resonant light that is linearly polarized with its $k$ vector parallel to the magnetic field so that it drives all $\sigma^+$ and $\sigma^-$ transitions. The power and frequency of the laser light are controlled with acousto-optic modulators (AOMs), which are in turn driven with radio-frequency pulse sequences programed by a field gate programmable array \cite{Pruttivarasin2015}.

We determine the ac Stark shifts of the $D_{5/2}$ state Zeeman sublevels, $\delta_{\mathrm{S}}(m_D)$, by measuring the difference between the $S_{1/2}\rightarrow D_{5/2}$ electric quadrupole transition frequencies with and without the off-resonant light. The transition frequencies are determined by probing the half-width at half maximum (HWHM) of the Rabi line shape around the transition centers, as described in \cite{Holliman2022}. We track each of the $S_{1/2}$, $m_S=\pm1/2\rightarrow D_{5/2}$, $m_D=\pm1/2$, $3/2$, $5/2$ transition frequencies with dedicated servos.

The ac Stark shift measurement pulse sequence starts with state detection ($\SI{0.5}{\milli\second}$) and Doppler cooling by driving the $S_{1/2}\leftrightarrow P_{1/2}\leftrightarrow D_{3/2}$ cycling transitions [see Fig. \ref{fig:sequence}(a)]. The ion is then prepared in either of the $m_S=\pm1/2$ Zeeman sublevels of the $S_{1/2}$ state via frequency-resolved $S_{1/2}\rightarrow D_{5/2}$ pumping \cite{Holliman2022}. After state preparation, a 250-$\mu$s-long $S_{1/2}\rightarrow D_{5/2}$ spectroscopy pulse is applied, during which the off-resonant light is either on, causing an ac Stark shift, or off, with no Stark effect. A final state detection determines if the ion is shelved in the $D_{5/2}$ state. The sequence above is repeated for positive and negative HWHM spectroscopy pulses for each of the six Zeeman transitions.

The pumping rate $\mathcal{R}(m_D)$ is measured by preparing the ion in a Zeeman sublevel of the $D_{5/2}$ state and probing the population remaining in the $D_{5/2}$ state after a variable-length off-resonant 802 nm pulse. The pumping rate pulse sequence starts with Doppler cooling, state detection, and state preparation, similar to the ac Stark shift sequence [see Fig. \ref{fig:sequence}(b)]. The ion is then shelved into a Zeeman sublevel of the $D_{5/2}$ state via a $S_{1/2}\rightarrow D_{5/2}$ pulse, followed by state detection to confirm shelving. The shelving pulse duration is optimized for shelving efficiency, which is limited to approximately $50\%$ by motional decoherence. Then the off-resonant light is turned on for a variable duration between 50 and 800 $\mu$s. Another state detection pulse follows to probe the population in the $D_{5/2}$ state. The pumping rate is determined by measuring the remaining $D_{5/2}$ state population as a function of the off-resonant light duration. The $D_{5/2}$ state population data are fit to a model (see Appendix \ref{sec:statistical_model} for details) that accounts for the different Rabi frequencies of all $D_{5/2}\rightarrow P_{3/2}$ Zeeman transitions as well as inelastic Raman scattering from the $P_{3/2}$ state to a different $D_{5/2}$ Zeeman sublevel \cite{Meir2020}.

The ac Stark shift and pumping rate measurements are interleaved to minimize systematic effects from slow (less than $0.1$ Hz) temporal drifts of the off-resonant light intensity and detuning.  We group ac Stark shift and pumping rate measurements into approximately 1-hour measurement blocks. The $D_{5/2}\rightarrow P_{3/2}$ transition frequency is measured before and after each block by preparing the ion in either of the $D_{5/2}$, $m_D=\pm5/2$ states and measuring the pumping rate to the $S_{1/2}$ and $D_{3/2}$ states as a function of the 802-nm laser frequency. We fit the spectrum with a Lorentzian function and take the transition center frequency as the average of the pumping rate peaks from the $m_D=\pm5/2$ states to cancel the linear Zeeman effect. Both the $D_{5/2}\rightarrow P_{3/2}$ transition frequency and the frequency of the off-resonant 802 nm light are tracked with a wavelength meter (High-Finesse WS8-10) to determine the detuning of the off-resonant light.

\section{Results}

\begin{figure}
    \centering
    \includegraphics[width=\figsize\linewidth]{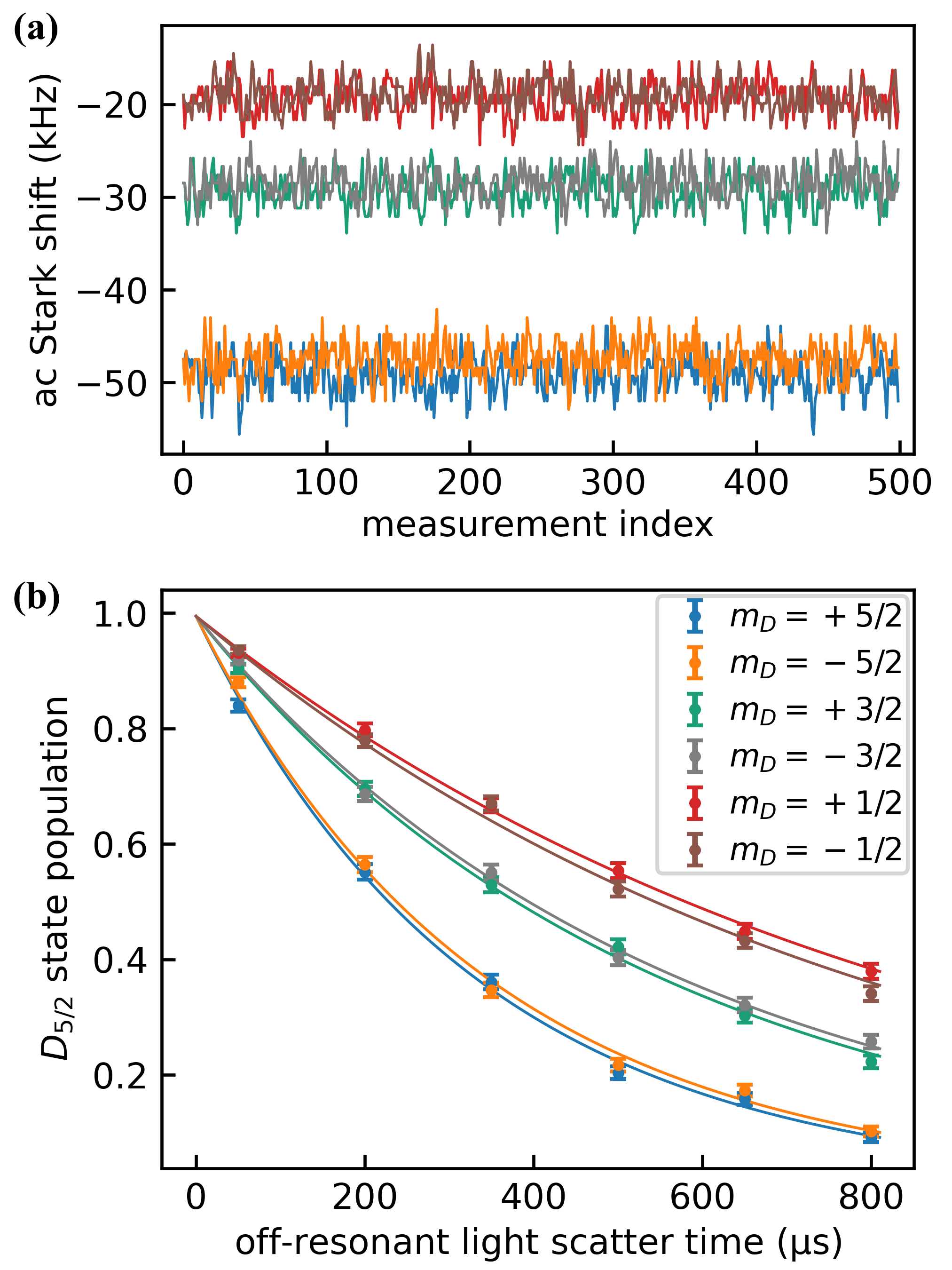}
    \caption{Data for a single measurement block.  (a) Measured ac Stark shift for each of the $D_{5/2}$ Zeeman sublevels. (b) $D_{5/2}$ state population as a function of the $D_{5/2}\rightarrow P_{3/2}$ off-resonant 802 nm pulse duration. Both plots use data from the same measurement block. The detuning of the off-resonant light to the $D_{5/2}\rightarrow P_{3/2}$ transition center is $\SI{-2998\pm3}{\mega\hertz}$, uncorrected for systematic effects. The legend is common to both plots.}
    \label{fig:datapoints}
\end{figure}

\begin{figure}
    \centering
    \includegraphics[width=\figsize\linewidth]{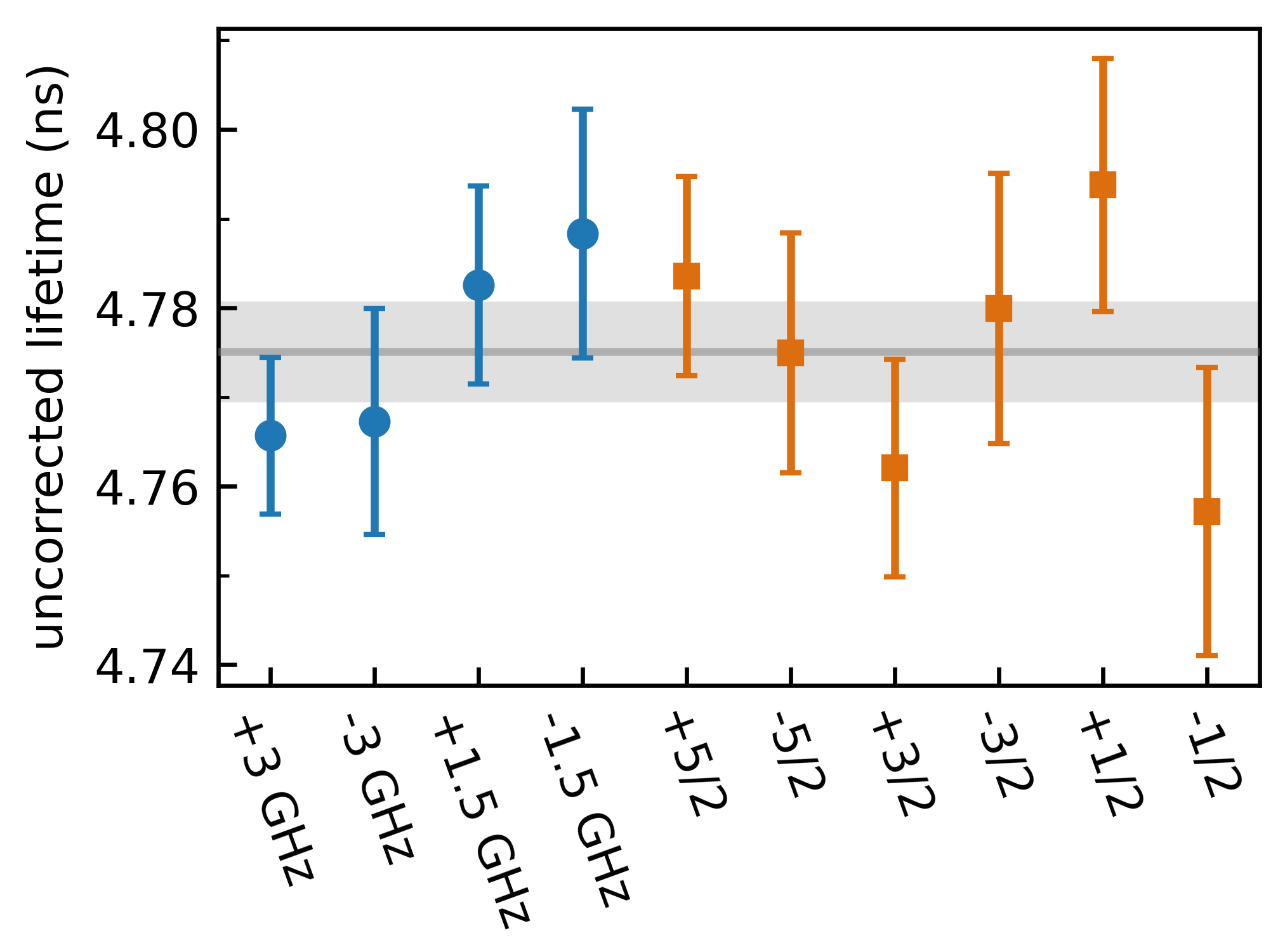}
    \caption{Uncorrected $P_{3/2}$ state lifetimes with the data points grouped by detunings (blue, circle) and $D_{5/2}$ Zeeman sublevels (orange, square). The total uncorrected lifetime and its uncertainty are shown as the gray region.}
    \label{fig:statistical}
\end{figure}

The results from one measurement block are shown in Fig. \ref{fig:datapoints}. For each $D_{5/2}$ Zeeman sublevel in a block, we calculate a $P_{3/2}$ state lifetime, uncorrected for systematic effects, from the measured ac Stark shift, pumping rate, and detuning using Eq. (\ref{eq:lifetime}). The uncorrected $P_{3/2}$ state lifetime, $4.775(6)$ ns, is calculated from an average of 246 measurement blocks. We also average subsets of the 246 measurement blocks with the same detunings or the same $D_{5/2}$ Zeeman sublevels (see Fig. \ref{fig:statistical}) to confirm that different detunings or Zeeman sublevels do not contribute to systematic shifts that are statistically significant.

The leading systematic uncertainty is due to the precision of the wavelength meter, which limits the uncertainty of $\Delta(m_D)$ and therefore the uncertainty of the lifetime [see Eq. (\ref{eq:lifetime})]. We use the wavemeter accuracy (10 MHz) as the systematic uncertainty in the detuning, which results in a 0.02-ns systematic uncertainty of the $P_{3/2}$ state lifetime.

We also consider possible optical power imbalance between the off-resonant light pulses during the ac Stark shift and pumping rate measurements. The optical power is actively stabilized before the AOM that switches the off-resonant light, but rf heating of the switching AOM may lead to different optical power for different pulse lengths, shifting the measured lifetime due to a shift in the Rabi frequency. We measure the effect of AOM heating on the ac Stark shift measurements by changing the AOM turn-on time to stabilize the off-resonant 802 nm power during the spectroscopy pulse and test the effect on the pumping rate measurements by removing data points of different off-resonant light pulse lengths from the analysis (see Appendix \ref{subsec:aom_thermal_effect}).  We check this effect by turning on the off-resonant light during state preparation [Fig. \ref{fig:sequence}(a), panel 3], which does not affect state preparation, but does allow us to vary the AOM turn-on time. We assign a 0.008(9)-ns systematic shift and uncertainty due to AOM heating for ac Stark shift measurements and 0.007-ns systematic uncertainty (no systematic shift) for pumping rate measurements.

The finite detection time also contributes to systematic effects due to the natural lifetime of the metastable $D_{5/2}$ state. We follow the method in \cite{Meir2020} to determine a detection error rate of $0.14\%$, which we use as a bound on the fractional uncertainty of the $P_{3/2}$ state lifetime.

\begin{figure}
    \centering
    \includegraphics[width=\figsize\linewidth]{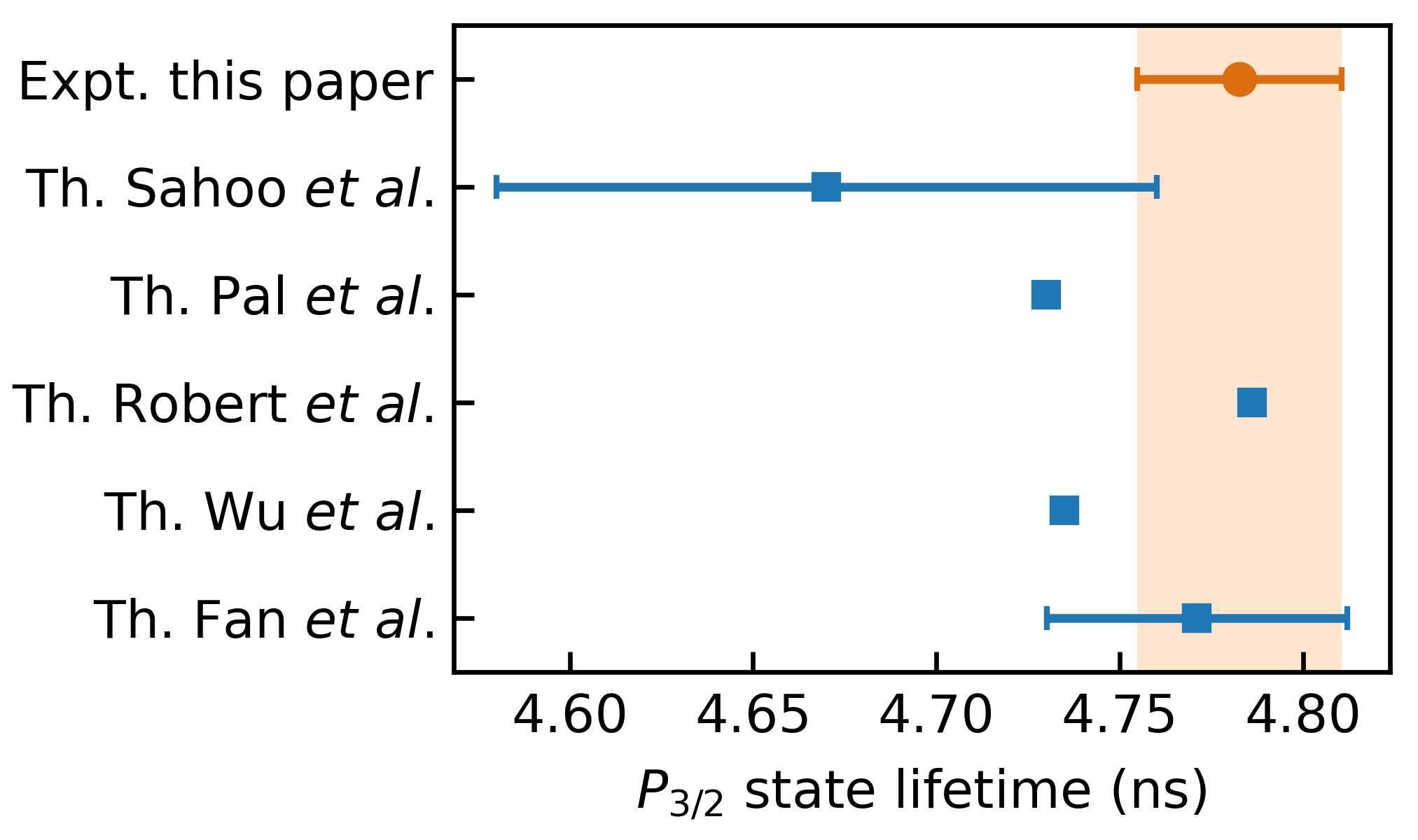}
    \caption{Comparison of the experimentally measured and theoretically calculated $P_{3/2}$ state lifetimes from Sahoo \textit{et al.} \cite{Sahoo2009}, Pal \textit{et al.} \cite{Pal2009}, Roberts \textit{et al.} \cite{Roberts2013a}, Wu \textit{et al.} \cite{Wu2016},  and Fan \textit{et al.} \cite{Fan2019a}.}
    \label{fig:theory}
\end{figure}

All shifts and uncertainties of systematic effects discussed above are shown in Table \ref{table:lifetime_summary}. More details about systematic effects are discussed in Appendix \ref{sec:systematic_effects}, including additional systematic effects that contribute less than $1$ ps to the measured lifetime. We add the systematic shifts linearly and the uncertainties in quadrature to get the final result of the $P_{3/2}$ state lifetime, $\tau=4.78(3)$ ns. The final result agrees with theoretical calculated lifetimes as seen in Fig. \ref{fig:theory}.

\begin{table}[t]
\caption{Summary of systematic shifts and uncertainties of the $P_{3/2}$ state lifetime. All units are in nanoseconds. The reported uncertainties represent one standard deviation. Known systematic effects with shifts and uncertainties smaller than 1 ps are not included.}
\label{table:lifetime_summary}
\begin{ruledtabular}
\begin{tabular}{lcc}
    Name & Shift & Uncertainty \\ 
    \hline \\ [-1.5ex]
    %Statistical value & 4.775 & 0.006 \\
    wavemeter accuracy & ... & 0.024 \\
    AOM heating (ac Stark shift) & 0.008 & 0.009 \\
    AOM heating (pumping rate) & ... & 0.007 \\
    finite detection time & ... & 0.007 \\
    %\hline \\ [-1.5ex]
    %Total & 4.78 & 0.03 \\
\end{tabular}
\end{ruledtabular}
\end{table}

With the measured branching fractions of the $P_{3/2}$ state \cite{Fan2019a}, the transition wavelengths reported in \cite{Holliman2020}, and the $P_{3/2}$ state measured in this work, we determine the reduced dipole matrix elements $\langle S_{1/2}||D||P_{3/2}\rangle=4.484(13)ea_0$, $\langle D_{5/2}||D||P_{3/2}\rangle=4.788(14)ea_0$, and $\langle D_{3/2}||D||P_{3/2}\rangle=1.513(11)ea_0$.

We thank Amar Vutha for helpful discussions.  This research was performed under the sponsorship of the NSF (Grants No.~PHY-1912665 and No.~PHY-2146555), the ONR (Grant No.~N00014-21-1-2597), and a NIST Precision Measurement Grant (No.~60NANB21D185).

\clearpage
\appendix
\begin{widetext}

\section{Model for the pumping rate measurement}
\label{sec:statistical_model}

In this appendix we describe the model used to determine the off-resonant 802 nm light pumping rate from the $D_{5/2}$ state population decay data [see Fig. \ref{fig:datapoints}(b)]. The model considers all Zeeman sublevels of the $P_{3/2}$ and the $D_{5/2}$ states to account for inelastic scattering between the $D_{5/2}$ state Zeeman sublevels.

The Rabi frequency of the $D_{5/2} (m_D) \rightarrow P_{3/2} (m_P)$ Zeeman transition with linearly-polarized light is \cite{James1998}

\begin{equation}
    \Omega(m_D, m_P, \theta)= \Omega_0\sqrt{2J'+1}
    \begin{pmatrix}
        5/2 & 3/2 & 1 \\
        m_D & -m_P & (m_P-m_D)
\end{pmatrix}f(\theta),
\end{equation}
where $\Omega_0=e|\vec{E_0}|\langle D_{5/2}||D||P_{3/2}\rangle/\hbar$, $J'=3/2$ is the total angular momentum of the upper state, $\theta$ is the angle between the polarization vector and the quantization axis, and the factor $f(\theta)$ is given by

\begin{equation}
    f(\theta)=
     \begin{cases}
       \sqrt{3/2}\sin{\theta} &\quad\text{for }\sigma^+\text{\ or\ }\sigma^-\text{\ transitions (}m_D-m_P=\pm1\text{)} \\
       \sqrt{3}\cos{\theta} &\quad\text{for }\pi\text{\ transitions (}m_D-m_P=0\text{)}.
     \end{cases}
\end{equation}

Zeeman transitions that the linearly polarized off-resonant 802-nm light can couple to when $\theta=90^\circ$ are shown in Fig. \ref{fig:supp_zeeman}.

\begin{figure}[h]
    \centering
    \includegraphics[width=0.4\linewidth]{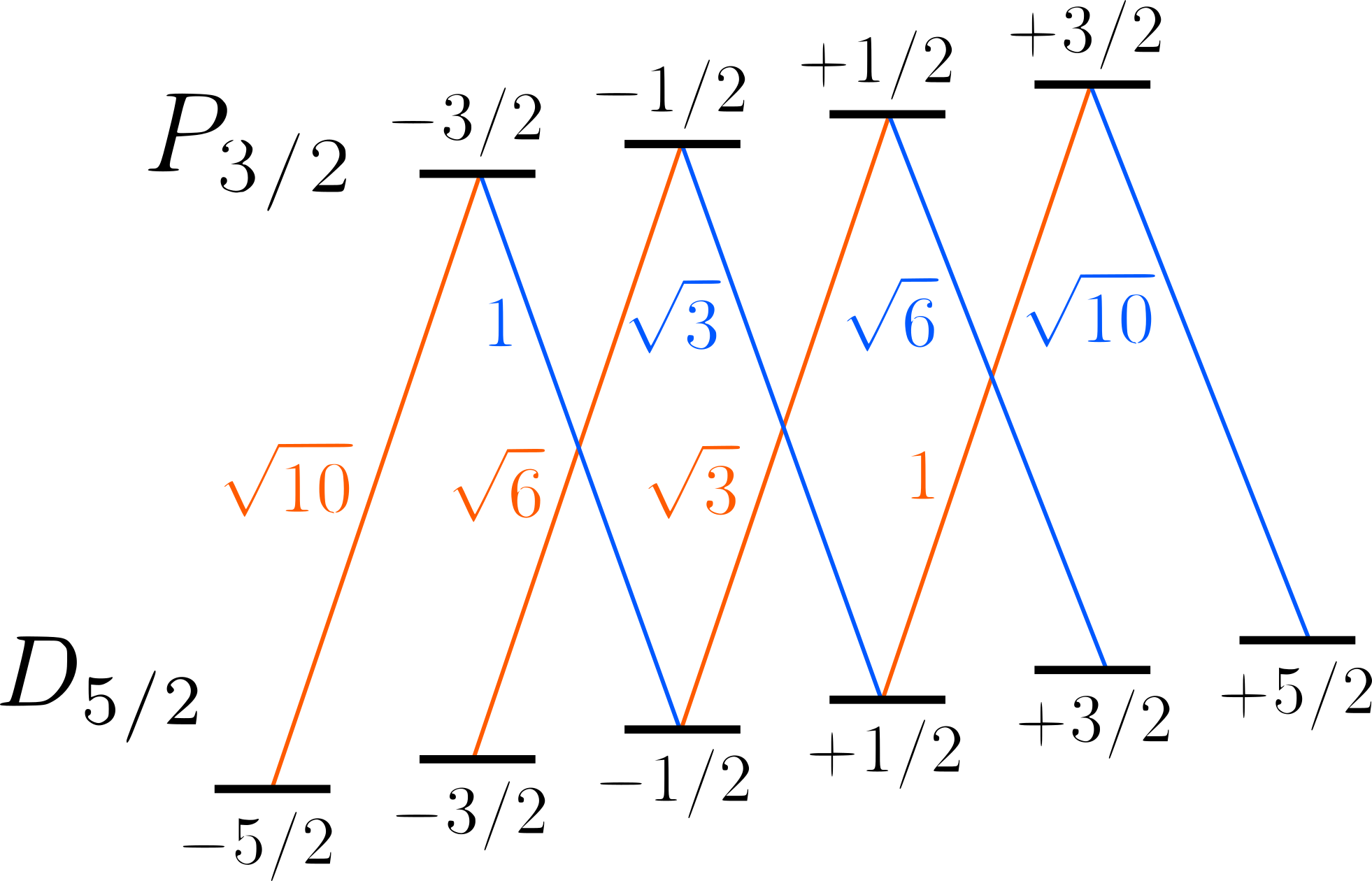}
    \caption{All allowed transitions between Zeeman sublevels labeled with the $m_D$ and $m_P$ quantum numbers in the $D_{5/2}$ and $P_{3/2}$ states with the off-resonant light propagating orthogonally to the magnetic field ($\theta=90^\circ$). The Rabi frequency ratios are labeled next to each of the transitions. Orange and blue denote the $\sigma^+$ and $\sigma^-$ transitions, respectively.}
    \label{fig:supp_zeeman}
\end{figure}

The scattering rate from $m_D$ to $m_P$ sublevels is

\begin{equation}
    R_{\mathrm{scatter}}(m_D, m_P) = 
    \Gamma\frac{\Omega(m_D, m_P, \theta=90^\circ)^2}{4\Delta(m_D, m_P)^2+2\Omega(m_D, m_P, \theta=90^\circ)^2+\Gamma^2}.
\label{eq:scattering_rate_mP}
\end{equation}
where $\Gamma=1/\tau$ is the spontaneous decay rate of the $P_{3/2}$ state.

With the approximations $\Gamma \ll \Delta(m_D, m_P)$ and $\Omega(m_D, m_P, \theta=90^\circ) \ll \Delta(m_D, m_P)$ (see Appendix \ref{subsec:line_shape})

\begin{equation}
    R_{\mathrm{scatter}}(m_D, m_P) = \Gamma\frac{\Omega(m_D, m_P, \theta=90^\circ)^2}{4\Delta(m_D, m_P)^2}.
\label{eq:scattering_rate_mP_approximated}
\end{equation}

The pumping rate from an $m_D$ sublevel to the $S_{1/2}$ and the $D_{3/2}$ states is

\begin{equation}
    \mathcal{R}(m_D) = (1-p) \sum_{m_P}R_{\mathrm{scatter}}(m_D, m_P),
\label{eq:pumping_rate_mP}
\end{equation}
where $p$ is the branching fraction from the $P_{3/2}$ state to the $D_{5/2}$ state.

Due to inelastic Raman scattering into other Zeeman sublevels of the $D_{5/2}$ state, the population in the $D_{5/2}$ state does not follow an exponential decay with the off-resonant light on. Instead of treating this deviation from an exponential decay as a systematic effect as in \cite{Meir2020}, we account for it in the analysis by fitting the solution to a set of differential equations that include the effect of inelastic Raman scattering to the data. The differential equations model the population in the $S_{1/2}$ and the $D_{3/2}$ states, as well as in all Zeeman sublevels of the $P_{3/2}$ and $D_{5/2}$ states. 

The $P_{3/2}$ state can decay to the $S_{1/2}$, $D_{3/2}$, or $D_{5/2}$ states with the branching fractions reported in \cite{Fan2019a}. The decay probability from an $m_P$ sublevel to an $m_D$ sublevel is given by:

\begin{equation}
    p(m_P, m_D)=p\sum_{q=-1}^{1}|\langle5/2,m_D;1, q|3/2, m_P\rangle|^2.
\end{equation}
where $\langle5/2,m_D;1, q|3/2, m_P\rangle$ is a Clebsch-Gordan coefficient.

The differential equations modeling the population in the states are
\begin{subequations}
\begin{align} 
\label{eq:diff_eqs_a}
\dot{P}_{P_{3/2}, m_P}(t) &= \sum_{m_D}P_{D_{5/2}, m_D}(t)R_{\mathrm{scatter}}(m_D, m_P) - P_{P_{3/2}, m_P}(t)/\tau, \\
\label{eq:diff_eqs_b}
\dot{P}_{D_{5/2}, m_D}(t) &= -P_{D_{5/2}, m_D}(t)\sum_{m_P}R_{\mathrm{scatter}}(m_D, m_P)+\sum_{m_P}p(m_P, m_D)P_{P_{3/2}, m_P}(t)/\tau-P_{D_{5/2}, m_D}(t)/\tau_{D_{5/2}}, \\
\label{eq:diff_eqs_c}
\dot{P}_{S_{1/2}}(t) + \dot{P}_{D_{3/2}}(t) &= \sum_{m_P}\left(1-p\right)P_{P_{3/2}, m_P}(t)/\tau+\sum_{m_D}P_{D_{5/2}, m_D}(t)/\tau_{D_{5/2}},
\end{align}
\end{subequations}
where $P_{P_{3/2}, m_P}(t)$ is the population in the $m_P$ sublevel, $P_{D_{5/2}, m_D}(t)$ is the population in the $m_D$ sublevel, $P_{S_{1/2}}(t)$ or $P_{D_{3/2}}(t)$ is the population in all Zeeman sublevels of the $S_{1/2}$ or $D_{3/2}$ state, $\tau$ is the $P_{3/2}$ state lifetime, and $\tau_{D_{5/2}}$ is the AOM-leakthrough limited $D_{5/2}$ state lifetime that we measured to be $\SI{0.27\pm0.03}{\second}$ (see Appendix \ref{subsec:finite_detect_time}).

The spontaneous decay rate of the $P_{3/2}$ state (theoretical value $\Gamma=2\pi\times33$ MHz \cite{Fan2019a}) is much greater than the off-resonant light pumping rates (less than $2\pi\times1.3$ kHz), so the population in the $m_D$ states can be regarded as time-independent for solving the differential equation for $P_{P_{3/2}, m_P}(t)$. With this approximation, Eq. (\ref{eq:diff_eqs_a}) is

\begin{equation}
    P_{P_{3/2}, m_P}(t)/\tau = \sum_{m_D}P_{D_{5/2}, m_D}(t)R_{\mathrm{scatter}}(m_D, m_P).
    \label{eq:equilibrium_mP}
\end{equation}

By plugging Eq. (\ref{eq:equilibrium_mP}) into Eqs. (\ref{eq:diff_eqs_b}) and (\ref{eq:diff_eqs_c}), the equations are no longer dependent on $P_{P_{3/2}, m_P}(t)$ and $\tau$,

\begin{align}
\dot{P}_{D_{5/2}, m_D}(t) &= -P_{D_{5/2}, m_D}(t)\sum_{m_P}R_{\mathrm{scatter}}(m_D, m_P)+\sum_{m_P}p(m_P, m_D)g_{m_P}(t)-P_{D_{5/2}, m_D}(t)/\tau_{D_{5/2}},\nonumber \\
\label{eq:diff_eqs_no_tau}
\dot{P}_{S_{1/2}}(t) + \dot{P}_{D_{3/2}}(t) &= \sum_{m_P}\left(1-p\right)g_{m_P}(t)+\sum_{m_D}P_{D_{5/2}, m_D}(t)/\tau_{D_{5/2}},
\end{align}
where $g_{m_P}(t) = P_{P_{3/2}, m_P}(t)/\tau$, with the initial conditions

\begin{align} 
P_{S_{1/2}}(0) &= 0, \nonumber\\
\label{eq:initial_conds}
P_{D_{3/2}}(0) &= 0, \\
P_{D_{5/2}, m_D}(0) &= \delta_{m_D,m_{D'}}, \nonumber
\end{align}
where $m_{D'}$ is the Zeeman sublevel of the $D_{5/2}$ state that the population is initially prepared in.

We fit a numerical solution to Eqs. (\ref{eq:diff_eqs_no_tau}) and (\ref{eq:initial_conds}) for the total $D_{5/2}$ state population $\sum_{m_D}P_{D_{5/2}, m_D}(t)$ to the data of each $m_D$ sublevel of a measurement block where $\Omega_0$ is the only fit parameter. The pumping rates from the $m_D$ Zeeman sublevels of the $D_{5/2}$ state are obtained from Eq. (\ref{eq:pumping_rate_mP}) using the fit results for the $\Omega_0$ values.

\section{Systematic Effects}
\label{sec:systematic_effects}

\subsection{Zeeman splittings}
\label{subsec:zeeman_splittings}

As shown in Fig. \ref{fig:supp_zeeman}, the off-resonant 802 nm light propagating orthogonally to the magnetic field only couples each of the $D_{5/2}$ state, $m_D=\pm3/2$ and $\pm5/2$ sublevels to one $P_{3/2}$ Zeeman state, so we can write

\begin{equation}
    \Delta(m_D=\pm3/2 \mathrm{\ or }\pm5/2, m_P) = \Delta(m_D) = \Delta(m_D, m_D - \mathrm{sgn}(m_D)),
\end{equation}
where $\mathrm{sgn}$ is the sign function.

From the $D_{5/2}$ state, $m_D=\pm1/2$ sublevels, the off-resonant light can drive both $\sigma^+$ and $\sigma^-$ transitions, and we approximate the detuning as the average of the detunings of the two Zeeman transitions

\begin{equation}
    \Delta(m_D=\pm1/2, m_P) \approx \Delta(m_D) = \frac{\Delta(m_D, m_D+1)+\Delta(m_D, m_D-1)}{2}.
\end{equation}

\end{widetext}

The maximum detuning error from either $\sigma^+$ or $\sigma^-$ transitions to the average detuning is 11 MHz ($0.4\%$ to $0.7\%$ fractional error). For states with opposite signs of $m_D$, the detuning errors are opposite, so the systematic shift cancels when we average the results from all Zeeman states. Therefore, we consider no systematic shift and uncertainty of the $P_{3/2}$ lifetime due to Zeeman splittings.

\subsection{Wavemeter accuracy}
\label{subsec:wavemeter_accuracy}
The detuning of the off-resonant $D_{5/2}\rightarrow P_{3/2}$ light is determined by the difference between frequencies of the off-resonant and the resonant 802-nm light measured using a HighFinesse WS8-10 wavemeter. Between measurement blocks, we calibrate the resonant 802-nm light frequency with $D_{5/2}\rightarrow P_{3/2}$ spectroscopy. The wavemeter has an accuracy of 10 MHz, which leads to a $0.5\%$ uncertainty in the detuning which translates to a 0.024-ns uncertainty in the measured $P_{3/2}$ state lifetime [see Eq. (\ref{eq:lifetime})].

\subsection{AOM thermal effect}
\label{subsec:aom_thermal_effect}

We use a single-pass AOM to control the off-resonant light driving the $D_{5/2}\rightarrow P_{3/2}$ transition. We apply approximately $\SI{100}{\milli\watt}$ of rf power to switch the state of the AOM. Due to thermal effects after applying the rf drive, it takes tens of microseconds before optical power is stabilized \cite{Meir2020}. Therefore, off-resonant light pulses with different pulse lengths may vary in average power. To reduce this effect, we send a 4-ms-long rf pulse at twice the nominal driving frequency to preheat the AOM controlling the off-resonant light before sending light to the ion using the nominal rf drive frequency. The preheat pulse thermalizes the AOM, but sends no light to the ion as the AOM driving frequency is out of both the AOM diffraction bandwidth and the fiber-coupling alignment bandwidth. Residual AOM thermal fluctuations are still possible due to a small heat load difference when it switches from the preheat pulse to the off-resonant pulse.

We test the residual AOM thermal shifts of the ac Stark shift measurements by turning on the off-resonant light 200 $\mu$s before the spectroscopy pulse driving the $S_{1/2}\rightarrow D_{5/2}$ transition and compare the results with the ac Stark shifts measured when the off-resonant light is turned on just before the spectroscopy pulse. With the 200-$\mu$s delay time, the optical power sent to the ion is stable for the entire duration of the spectroscopy pulse. We measure that the ac Stark shift with the 200-$\mu$s delay time is $0.16(18)\%$ larger than the ac Stark shift without, and thus the systematic shift of the $P_{3/2}$ state lifetime is 0.008(9) ns.

\begin{figure}
    \centering
    \includegraphics[width=\linewidth]{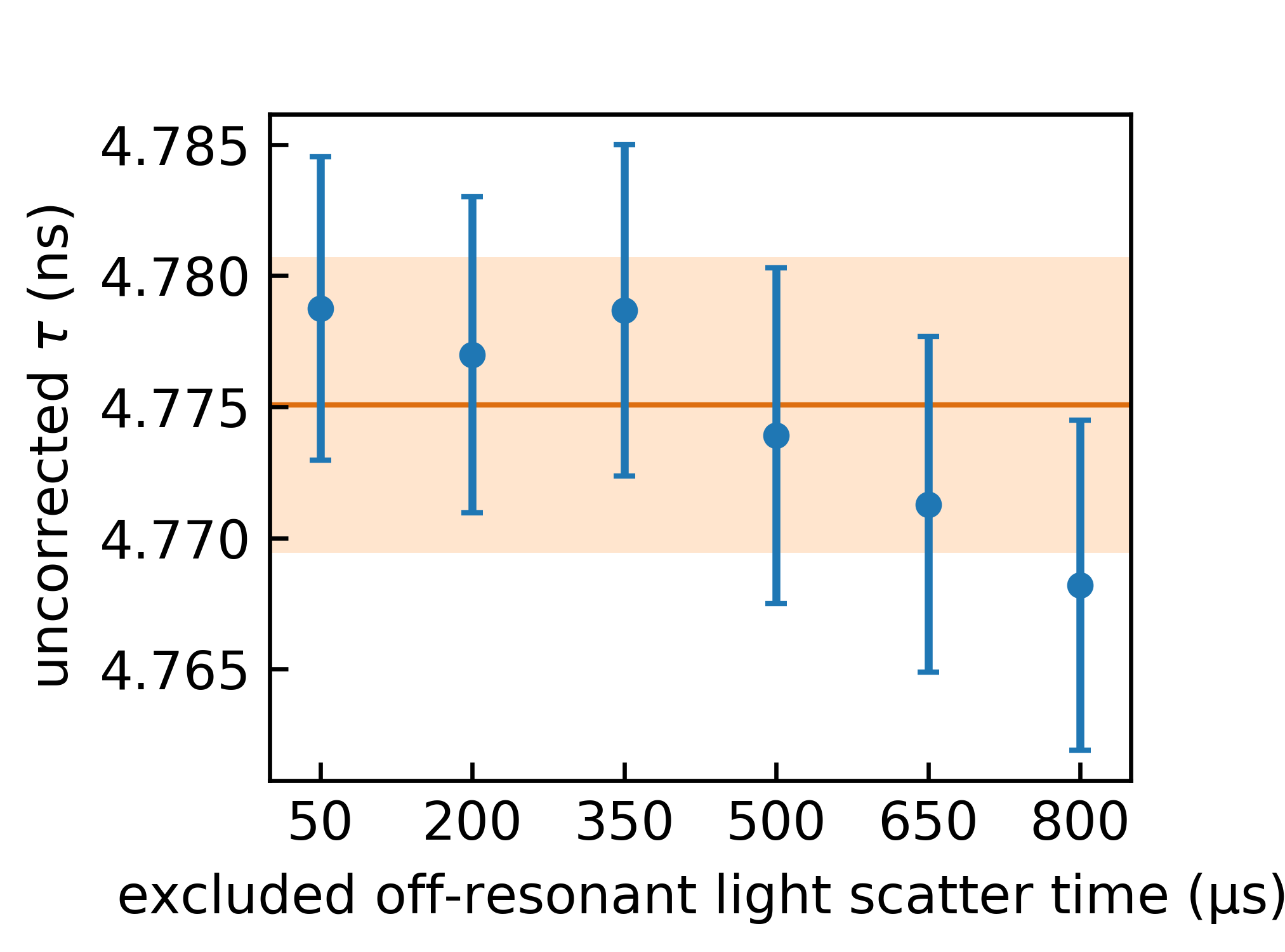}
    \caption{The uncorrected $P_{3/2}$ state lifetimes excluding data of corresponding off-resonant light scatter times are shown in blue. The uncorrected lifetime and uncertainty with all data are shown in orange. We take the systematic uncertainty due to AOM heating for the pumping rate measurements as the shift of the uncorrected lifetime with the 800-$\mu$s scatter time data removed, where the shift is the largest among removal of the six scatter times.}
    \label{fig:supp_datapoint_removal}
\end{figure}

For the pumping rate measurements, we anticipate the effect due to AOM heating to be more significant for data points with short durations, so we test the shift by removing each of the six off-resonant light durations from the analysis, as shown in Fig. \ref{fig:supp_datapoint_removal}. For example, the 50-$\mu$s data point in Fig. \ref{fig:supp_datapoint_removal} is the uncorrected lifetime from a fit [see Fig. \ref{fig:datapoints}(b)] that excludes the corresponding off-resonant light scatter time in the pumping rate measurement. The systematic uncertainty (0.007 ns) is determined as the maximum shift of the uncorrected lifetime with each of the scatter time data excluded.

\subsection{Finite detection time}
\label{subsec:finite_detect_time}

During state detections, both $S_{1/2}\rightarrow P_{1/2}$ and $D_{3/2}\rightarrow P_{1/2}$ transitions are driven and the $P_{1/2}\rightarrow S_{1/2}$ photons from the ion are collected. If the ion is in the $S_{1/2}$ or the $D_{3/2}$ state, on average we collect $N_\mathrm{b}=24$ photons during a state detection, which is referred to as a bright event. If the ion is in the $D_{5/2}$ state, an average of $N_\mathrm{d}=0.25$ photon is collected, which is referred to as a dark event. If the ion is in the $D_{5/2}$ state at the beginning of a state detection, decays of the $D_{5/2}$ state during the state detection may result in a bright event. We estimate the probability of such detection error following \cite{Meir2020},

\begin{equation}
    p_{\mathrm{error}} = 1 - e^{[t_{\mathrm{det}}(N_\mathrm{b}-N_\mathrm{t})/(N_\mathrm{b}-N_\mathrm{d})]/\tau_{D_{5/2}}},
\end{equation}
where $t_{\mathrm{det}}=0.5$ ms is the state detection time, $N_\mathrm{t}=5.5$ is the selected state detection threshold between dark and bright events, and $\tau_{D_{5/2}}$ is the lifetime of the $D_{5/2}$ state.

We measure the lifetime of the $D_{5/2}$ state to be 0.27(3) s, which is less than the theoretically calculated 0.303(4) s \cite{Pal2009} likely due to leakthrough light from the AOM that controls the resonant 802-nm light. The finite state detection time results in a dark state detection error rate of $0.14\%$, which we use as an upper bound for the fractional systematic uncertainty of the $P_{3/2}$ lifetime, corresponding to an absolute uncertainty of 0.007 ns.

\subsection{Off-resonant 802 nm light polarization error}
\label{sec:polar_error}

As the Rabi frequencies for Zeeman transitions depend on the polarization angle with the magnetic field, imperfect laser alignment and polarization may shift the measured pumping rates. From ratios between ac Stark shifts of different $D_{5/2}$ Zeeman sublevels, we measure that the angle between the polarization and the magnetic field is 88(2)$^\circ$, and residual circularly polarized light shifts the Rabi frequency ratio of $\sigma^+$ and $\sigma^-$ Zeeman transitions from unity to $\Omega_{\sigma^+}/\Omega_{\sigma^-}=0.986(5)$.

We account for the polarization error by fitting to the pumping rate model in Appendix \ref{sec:statistical_model} that includes the above polarization error. The systematic uncertainties with the linear polarization error and the residual circular polarization are $\SI{7e-4}{\nano\second}$ and $\SI{1.1e-4}{\nano\second}$ respectively.

\subsection{$P_{3/2}$ state branching fraction uncertainty}

The uncertainty of the $P_{3/2}$ state branching fraction to the $D_{5/2}$ state, $p=\SI{0.10759\pm0.00010}{}$ measured in \cite{Fan2019a} contributes to $\SI{5e-4}{\nano\second}$ uncertainty following Eq. (\ref{eq:lifetime}).

\subsection{Line shape}
\label{subsec:line_shape}

In Eqs. (\ref{eq:ac_stark_shift}) and (\ref{eq:pumping}), the pumping rate and the ac Stark shift were calculated without considering the linewidth of the $D_{5/2}\rightarrow P_{3/2}$ transition. With the Lorentzian line shape taken into account, the maximum fractional shift of the measured lifetime is \cite{Meir2020}

\begin{equation}
    \epsilon(m_D, m_P)\approx\frac{\Omega(m_D, m_P)^2/2+(1/\tau_{P_{3/2}}^2)/4}{\Delta(m_D, m_P)^2}.
\end{equation}
The associated systematic uncertainty is $\SI{5e-4}{\nano\second}$.

\subsection{Coupling to other dipole transitions}

We also account for the interaction of the probe light with the $S_{1/2}\rightarrow P_{1/2}$ and $S_{1/2}\rightarrow P_{3/2}$ transitions. The small but finite $S_{1/2}$ state ac Stark shift due to off-resonant coupling with these transitions is considered as a part of the $D_{5/2}$ state ac Stark shift in the analysis.

We calculate the ac Stark shift of the $S_{1/2}$ state to be less than $0.007\%$ of the ac Stark shift of the $D_{5/2}$ state. We use this fractional shift as an upper bound of the fractional systematic uncertainty due to couplings to other transitions, which gives a systematic uncertainty of $\SI{4e-4}{\nano\second}$. Moreover, the ac Stark shift of the $D_{5/2}$ state changes sign for red and blue detuned off-resonant light, but the ac Stark shift of the $S_{1/2}$ does not change. Therefore we anticipate that this systematic effect shifts the measured lifetime in opposite directions for detunings of opposite signs, which we do not observe up to the statistical uncertainty. All other transitions connecting to either the $D_{5/2}$ or the $S_{1/2}$ state are even further detuned and contribute negligibly to the ac Stark shifts.

\subsection{Raman transition}

The off-resonant light couples to all $\sigma^+$ and $\sigma^-$ transitions, allowing off-resonant Zeeman transitions between different Zeeman sublevels of the $D_{5/2}$ state with the selection rule $\Delta m_D=\pm2$. Following the treatment in \cite{Meir2020}, the maximum Rabi frequency for the Raman transition is $2\pi\times\SI{89}{\kilo\hertz}$ and the corresponding systematic uncertainty of the $P_{3/2}$ lifetime is $\SI{2e-4}{\nano\second}$.

\subsection{Detection fidelity}

The collected photons in bright and dark state detections follow Poisson distributions around their average, except for the $D_{5/2}$ decay events which were considered in Appendix \ref{subsec:finite_detect_time}. Given the average bright and dark counts and the threshold, the probability for a bright or dark event to be misidentified based on Poisson distributions is $\SI{3e-6}{}$ or $\SI{3e-7}{}$. The maximum error probability, $\SI{3e-6}{}$, is used to bound the fractional systematic uncertainty of the $P_{3/2}$ lifetime, corresponding to an absolute uncertainty of $\SI{1.5e-5}{\nano\second}$.

\subsection{Ion motion}

Following \cite{Meir2020}, the systematic shift of the $P_{3/2}$ lifetime due to ion motion is $\epsilon=\beta^2$, where $\beta=2\pi x \omega/(\lambda\Delta)$ is the modulation index of the motion, $x$ is the ion's motional amplitude at oscillation frequency $\omega$, and $\lambda$ is wavelength of the off-resonant light. We determine $\beta_{\mathrm{micromotion}}<0.0019$ and $\beta_{\mathrm{thermal}}\sim0.00014$, which give $\SI{1.8e-5}{\nano\second}$ and $\SI{1e-7}{\nano\second}$ for systematic uncertainties associated with micromotion and thermal motion of the ion.

\bibliographystyle{apsrev4-2}

\end{document}